\documentclass{nature}

\usepackage{lineno}
\usepackage{braket}
\usepackage{graphicx}
\usepackage{amsmath}
\usepackage{amssymb}
\usepackage[T1]{fontenc}

\usepackage{color}
\newcommand{\means}[1]{\langle#1\rangle}
\usepackage{bm}
\title{The range of non-Kitaev terms and fractional particles in $\alpha$-RuCl$_{3}$}

\author{Yiping Wang,$^{1}$ Gavin B. Osterhoudt,$^{1}$ Yao Tian,$^{2}$ Paige Lampen-Kelley,$^{3}$ Arnab Banerjee, $^{10}$ Thomas Goldstein,$^{5}$ Jun Yan,$^{5}$ Johannes Knolle,$^{6}$ Huiwen Ji,$^{7}$, Robert J. Cava,$^{7}$ Joji Nasu,$^{8}$ Yukitoshi Motome,$^{9}$ Stephen E. Nagler,$^{4}$ David Mandrus,$^{3}$ Kenneth S. Burch$^{1}$}

\begin{document}
\maketitle

\begin{affiliations}
 \item Department of Physics, Boston College, 140 Commonwealth Avenue, Chestnut Hill, MA 02467, USA
 \item SICK Product Center Asia Pte. Ltd., 8 Admiralty Street, Singapore 757438
 \item Department of Materials Science and Engineering, University of Tennessee, Knoxville, TN 37821, USA
 \item Neutron Scattering Division, Oak Ridge National Laboratory, Oak Ridge, TN 37831, USA
 \item Department of Physics, University of Massachusetts, Amherst, Massachusetts 01003, USA
 \item Blackett Laboratory, Imperial College London, London SW7 2AZ, United Kingdom
 \item Department of Chemistry, Princeton University, Princeton, NJ 08540, USA 
 \item Department of Physics, Yokohama National University, Hodogaya, Yokohama 240-8501, Japan
 \item Department of Applied Physics, University of Tokyo, Bunkyo, Tokyo 113-8656, Japan
 \item Department of Physics and Astronomy, Purdue University, West Lafayette, IN 47906, USA
\end{affiliations}

\begin{abstract}
Significant efforts have focused on the magnetic excitations of relativistic Mott insulators, predicted to realize the Kitaev quantum spin liquid (QSL). This exactly solvable model involves a highly entangled state resulting from bond-dependent Ising interactions that produce excitations which are non-local in terms of spin flips. A key challenge in real materials is identifying the relative size of the non-Kitaev terms and their role in the emergence or suppression of fractional excitations. Here, we identify the energy and temperature boundaries of non-Kitaev interactions by direct comparison of the Raman susceptibility of $\alpha$-RuCl$_{3}$ with quantum Monte Carlo (QMC) results for the Kitaev QSLs. Moreover, we further confirm the fractional nature of the magnetic excitations, which is given by creating a pair of fermionic quasiparticles. Interestingly, this fermionic response remains valid in the non-Kitaev range. Our results and focus on the use of the Raman susceptibility  provide a stringent new test for future theoretical and experimental studies of QSLs. 
\end{abstract}


Exotic excitations with fractional quantum numbers are a key characteristic of QSLs\cite{Wen2016,Nussinov2013ROM,Burch2018Nat,Takahashi2019PRX}, which result from the long range entanglement of these non-trivial topological phases\cite{Lee2008Science,Savary2017,Han2012Nat}. Originating from frustrated magnetic interactions, the fractional nature inspires an overarching goal of studying QSLs, realizing topological quantum computing immune to decoherence, with high operating temperatures from large exchange interactions\cite{Kitagawa2018Nat,Jackeli2009PRL}. The last decade has seen great progress towards identifying the fractional excitations of QSLs\cite{Catuneanu2018npj,sandilands2015PRL,Banerjee2016NatMat,wulferding2010PRB,Kitaev2006,Vojta2016,Nasu2014,Becker2015PRB,Seifert2018}. Attention has focused on relativistic Mott insulators that are close to the exactly solvable Kitaev model with a  QSL ground state. In materials such as A$_2$IrO$_{3}$ (A = Cu, Li or Na)\cite{singh_NaIrO3_2010PRB,Kitagawa2018Nat,Abramchuk2017JACS,Matern2017JSM,Comin2012PRL,modic2014NatComm,Takahashi2019PRX} and $\alpha$-RuCl$_{3}$\cite{PLumb2014,Sandilands2016PRB,Glamazda2017,Zheng2017PRL}, the large spin-orbit coupling and Coulomb repulsion result in j$_{eff}=1/2$ moments on a honeycomb lattice\cite{Nussinov2013ROM,Hae-Young2016PRB,Winter2017NatComm,Do2017Natphys,Sandilands_SOE_2016PRB,Jackeli2009PRL,Zhou2016PRB,Koitzsch_EELS_2016PRL,Nasu2016,Yan2011}.According to the pure Kiteav model, in these materials spin flips could produce Z$_2$ gauge  fluxes  and  dispersive Majorana fermions.\cite{Kitaev2006,Knolle2014}.

In these candidate materials, as with other QSL candidates, the presence of additional symmetry allowed terms (Heisenberg and bond-dependent off-diagonal interactions in this case), produces long range magnetic order.\cite{Winter2017NatComm,Catuneanu2018npj,Chaloupka2010PRL,Catuneanu2018npj} Despite extensive studies and evidence for fractional particles\cite{sandilands2015PRL,Nasu2016,Kasahara2018Nat,Banerjee2017Science}, the relative size of the non-Kitaev terms and the range over which they are relevant remains controversial\cite{Chaloupka2010PRL,Rousochatzakis2019}. In $\alpha$-RuCl$_{3}$, these non-Kitaev terms lead to a magnetically ordered phase below 7~K, which could be destroyed by an in-plane magnetic field\cite{Leahy2017PRL,Banerjee2018npj,Yadav2016,WangLoidl2017PRL,Jansa2018NatPhys}. The exact nature of the field induced QSL state remains unclear\cite{Zheng2017PRL,Kasahara2018Nat,Banerjee2018npj} as the zero field Hamiltonian is still unresolved. In particular, non-Kitaev interactions dominant energy and temperature ranges have not yet been experimentally established. Additionally, there is a need to determine if excitations in these ranges maintain their fractional nature. 

Raman scattering is a powerful probe of magnetic materials, revealing the presence of long range order, symmetry and statistics of the excitations, as well as the strength and nature of the exchange, even in single 2D atomic layers\cite{Reiter1976PRB,fleury1968scattering,Kim2019NatComm,Zhou2018JPCS,Misochko1996,kuroePRB97,Nakamura2015JPSJ,GretarssonTacon2016PRL,Burch2018Nat}. Indeed, Raman scattering was the first to reveal the continuum from magnetic excitations in $\alpha$-RuCl$_{3}$\cite{sandilands2015PRL}. However, a careful study of the Kitaev term's temperature and energy dependence is still a challenge, as one requires a very high temperature and energy resolution to show the spectral change and directly compare the spectra with theoretical calculations\cite{Winter2018PRL}. Previously, Raman efforts relied on spectral integration over a certain energy range which averaged out the energy dependence of the excitations, and, the low scattering intensity made it difficult to directly compare the spectra with theoretical calculations from the exact Kitaev model. As such the role of the non-Kitaev terms, and their size, could be identified in previous efforts. Furthermore, demonstration of the fractional nature relied on the  integrated Raman intensity and thus required  subtraction of a bosonic background, without justification. This approach also meant fitting the data with an average energy in the fermi function, further limiting the ability to uncover if the non-Kitaev terms affected the statistical response of the excitations.\cite{Glamazda2017,Nasu2016}.

Here, we overcome all these previous limitations with new Raman spectra with dramatically improved signal levels, high temperature and energy resolution. Firstly, having improved the optics, our Raman measurements now obtain a signal level 18 times larger than before\cite{sandilands2015PRL} (see supplemental). This high signal level provides enough anti-Stokes response to ensure the temperature is correct and allows us to directly extract the Raman susceptibility by taking the difference between Stokes and anti-Stokes intensities, after the dark counts have already been removed. In this way, the role of the non-Kitaev terms is revealed via a direct comparison of the full temperature and energy dependence of the $\alpha$-RuCl$_{3}$ Raman susceptibility with a QMC calculation for the pure Kitaev model. Furthermore, we provide compelling evidence for the fermionic nature of the excitations without the need to subtract a bosonic background. Our results show that the Raman susceptibility of $\alpha$-RuCl$_{3}$ is consistent with QMC calculations at higher temperatures and energies ($>$40K \& $>$6meV). The deviation between them in the low temperature and energy range ($<$40K \& $<$6meV) results from the non-Kitaev terms. Via these temperature and energy boundaries, we directly measure the ratio of $J_K$ and $\Gamma$ interactions in the Hamiltonian. Moreover, the high temperature(>150K) difference between the Raman susceptibility and the QMC indicates the presence of quasi-elastic scattering (QES) induced by thermal fluctuations in the system, commonly observed in frustrated systems\cite{Han2012Nat,sandilands2015PRL, Glamazda2017}. With our enhanced signal the anti-Stokes spectra for all temperatures can be compared with the Stokes response to prove the sample is in detailed balance without laser heating (see Fig.~\ref{fig:Fig2}d). To further explore the effect of non-Kitaev interactions, we fit the Raman susceptibility with a Fermi function containing half the measured energy. The very good overlap shows the excitations are governed by Fermi statistics even beyond the Kitaev dominant range.\cite{Nasu2015PRB} We also checked that the susceptibility integration is governed by a Fermi function with half energy, which further confirms each fractional particle holds one half of the scattering energy in both Kitaev and non-Kitaev dominant regimes. Interestingly this is revealed without the need to subtract the bosonic background.


In inelastic light scattering, the measured intensity is determined by symmetry, Fermi's golden rule, and from the fluctuation-dissipation theorem, is proportional to the Raman susceptibility ($\text{Im}(\chi [\omega,T])$) times a Bose function\cite{devereaux2007RoM,Butcher1965gw}. In magnets this can produce peaks from single magnons, broad features reflecting the two-magnon joint density of states (JDos), or QES from thermal fluctuations\cite{sandilands2015PRL,wulferding2010PRB}. For the Kitaev QSL, Raman predominantly excites pairs of fractional particles in the energy range considered here ($\approx 0.5J_K<\hbar\omega<\approx 2J_K$), leading to the energy loss ($I_{S}[\omega,T]$) and gain ($I_{aS}[\omega,T]$) intensities\cite{knolle2014raman,Nasu2016}:

\begin{align*}
I_S[\omega,T] = \text{Im}(\chi [\omega,T]) (n_B[\omega,T]+1) = JDos[\omega,T](1-n_F[\omega/2,T])^2\\
I_{aS}[\omega,T] = \text{Im}(\chi [\omega,T]) (n_B[\omega,T]) = JDos[\omega,T](n_F[\omega/2,T])^2
\end{align*}

where $n_{B/F}[\omega,T]$ are the Bose/Fermi distributions and $JDos[\omega,T]$ is approximately given by the JDos from the fractional particles. For responses from non-fractional excitations, for example phonons, we expect an additional term to be added to the susceptibility, without contributions from the Fermi function. 

\begin{figure}
    \centering
    \includegraphics[width=0.7\textwidth]{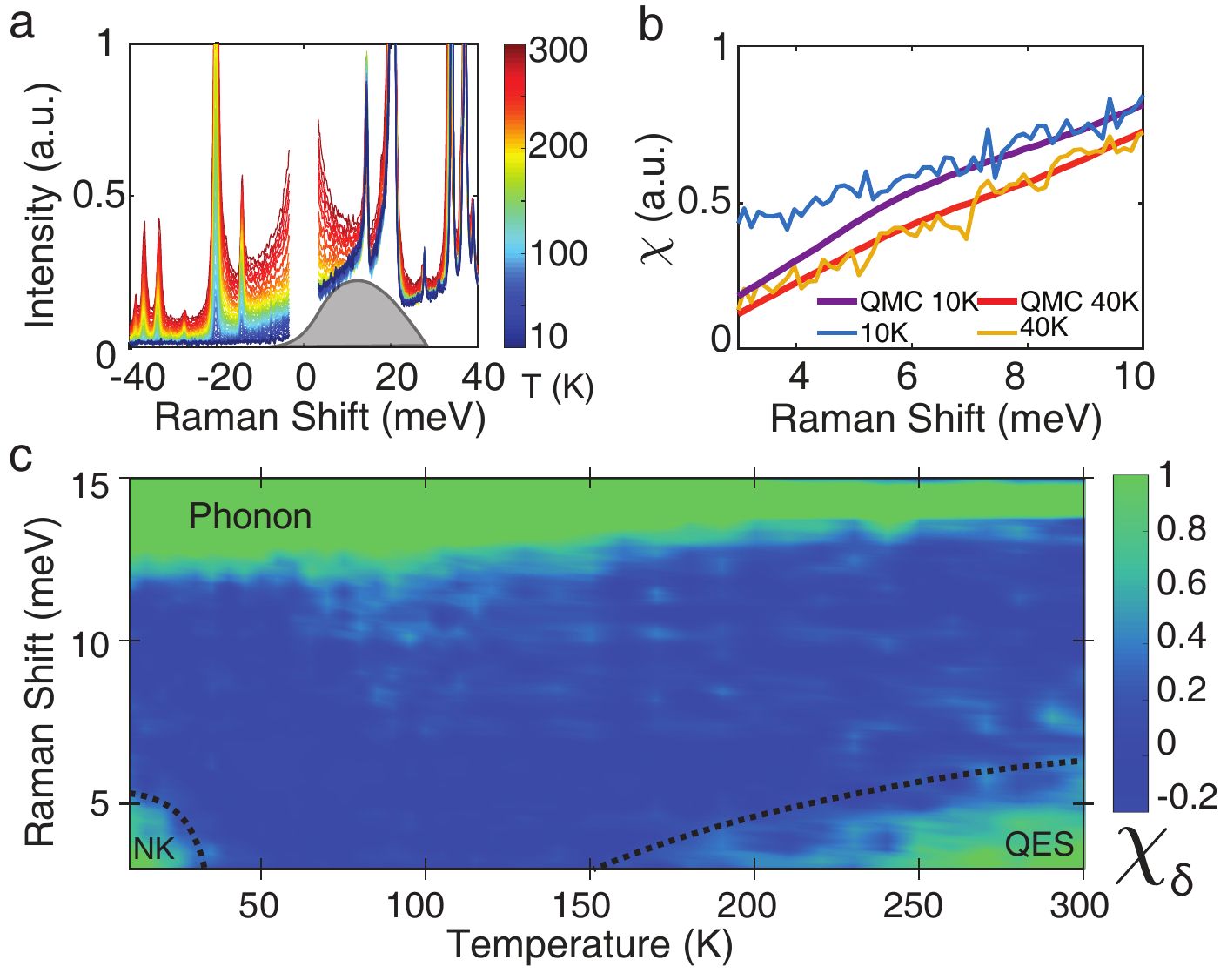}
    \caption{\textbf{Effects of non-Kitaev Terms} 
    (\textbf{a}) Temperature dependent Raman intensity of $\alpha$-RuCl$_{3}$ in XY polarization. Both Stokes and anti-Stokes data are collected from 10 K to 300 K with 5 K steps below 120 K and 10 K steps above. the gray shade is indicates the magnetic continuum excitation. (\textbf{b}) The measuredRaman susceptibility in XY polarization of $\alpha$-RuCl$_{3}$ at 10 K (blue line) compared with the calculated result of the pure Kitaev limit (purple line) at the same temperature. The enhanced signal at low energies results from the non-Kitaev interactions in the system. By 40 K there is nearly perfect agreement between the Raman data (yellow line) and the QMC calculation (red line), indicating the non-Kitaev terms are not relevant in this energy and temperature range. (\textbf{c}) The temperature and energy dependent map of $\chi_{\delta}$($\chi_{\delta} = \chi_{measured}-\chi_{QMC}$). $\chi_{\delta}$ at low temperature and low energy range shows the temperature and energy boundary of non-Kitaev (NK) interactions in the system. $\chi_{\delta}$ at the high temperature and low energy range indicates the quasi-elastic scattering (QES) in the system. }
    \label{fig:Fig1}
\end{figure}

\begin{figure}
\includegraphics[width=0.8\textwidth]{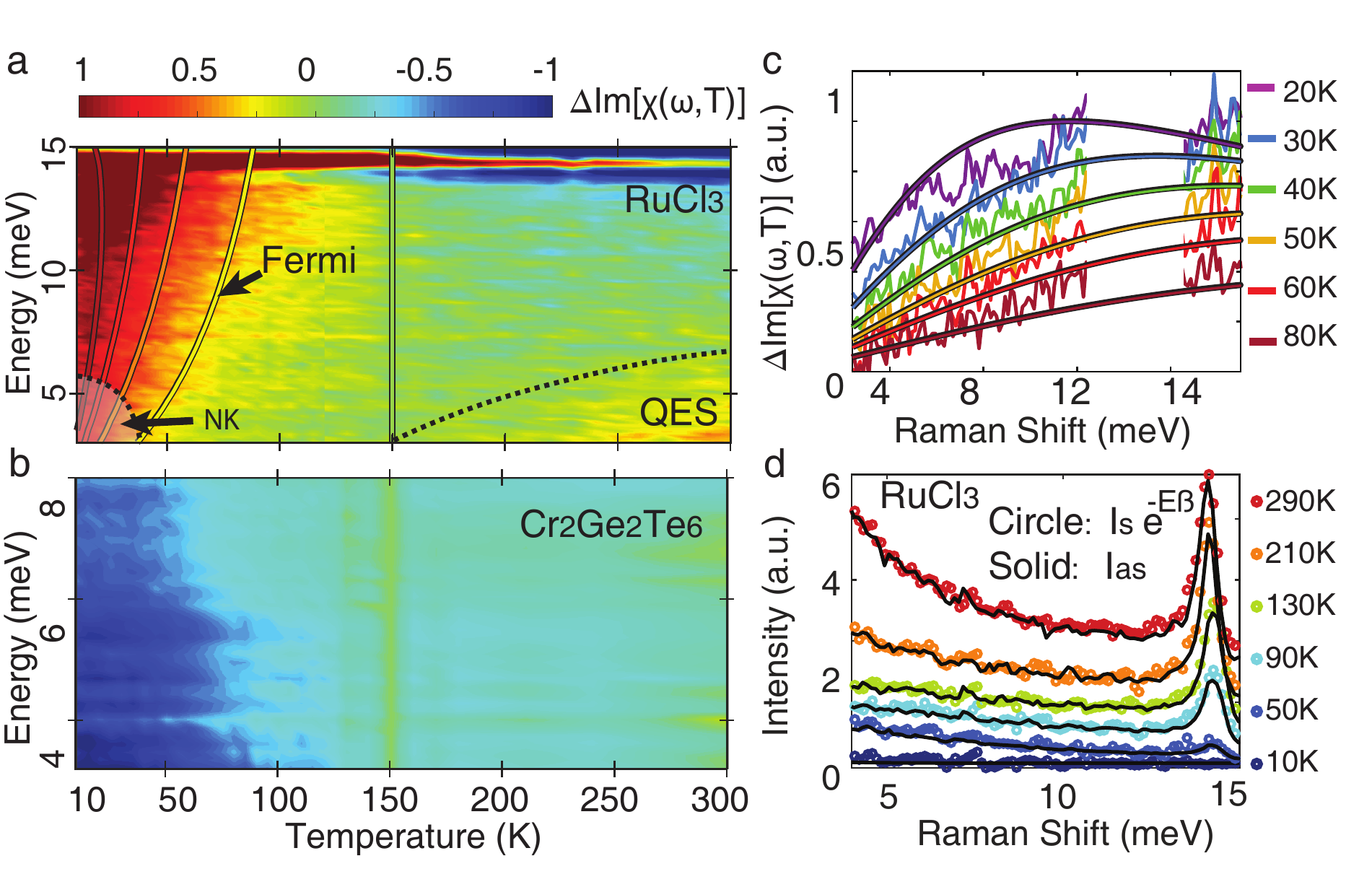}
    \caption{\textbf{The normalized Raman susceptibility and detailed balance}
 (\textbf{a}) Raman susceptibility of RuCl$_{3}$, $\Delta \text{Im}[\chi(\omega,T)] = \text{Im}[\chi(\omega,T)]-\text{Im}[\chi(\omega,150~\text{K})]$). The curves with black outlines are the contour plots of the Fermi function ($\Delta n_{F}(\omega/2,T)=n_{F}(\omega/2,150)-n_{F}(\omega/2,T)$). Both data and the prediction are normalized to their maximum values. The agreement between the two confirms that Raman creates magnetic excitations that are made of pairs of fermions. The upturn of the Raman intensity in the high temperature and low energy range results from thermal fluctuations of the magnetism (quasi-elastic scattering). (\textbf{b}) Raman susceptibility of a similar magnet, Cr$_{2}$Ge$_{2}$Te$_{6}$, where, opposite to $\alpha$-RuCl$_{3}$, $\Delta Im[\chi(\omega,T)]$ is negative and does not match $n_{F}(\omega/2,T)$. (\textbf{c}) Comparison of $n_{F}(\omega/2,T)$ and $\Delta \text{Im}[\chi(\omega,T)]$ of RuCl$_3$ at fixed temperatures. The agreement further confirms the excitations are fermionic. (\textbf{d}) The excellent agreement between Stokes and anti-Stokes spectra of $\alpha$-RuCl$_{3}$ when normalized by the Boltzmann factor demonstrates the absence of laser heating.}
    \label{fig:Fig2}
\end{figure}

As shown in Fig.~\ref{fig:Fig1}a, we collected both the Stokes and anti-Stokes spectra of bulk $\alpha$-RuCl$_{3}$ from 10 K to 300 K. Our Rayleigh scattering half width is 2.3 meV, enabling measurement down into the low energy regime. The temperature dependent spectra show a clear magnetic excitation continuum (2.3meV $\sim$ 10 meV) below the first phonon, which mostly results from the Kitaev interaction and is consistent with previous predictions and measurements\cite{sandilands2015PRL,Banerjee2017Science,Glamazda2017,Nasu2016}. Since the measured Raman intensity contains a Bose factor, it is best to investigate the Raman susceptibility $\text{Im}[\chi[\omega,T]]$\cite{devereaux2007RoM,wulferding2010PRB,Nakamura2015JPSJ,Misochko1996}. Using our new anti-Stokes spectra, we directly determine the susceptibility from the difference between the Stokes and anti-Stokes intensities ($I_{S}[\omega,T]-I_{aS}[\omega,T]=\text{Im}[\chi[\omega,T]]$). This new data set, combined with  minimizing the rise in temperature due to the laser, reveals the regimes in which non-Kitaev terms are relevant. Specifically, Fig.~\ref{fig:Fig1}b shows the comparison of the QMC results for the pure Kitaev limit and the Raman susceptibility at 10 and 40 K. While excellent agreement is seen at 40 K, the data at 10 K only matches the model between 6 to 10 meV. Noting that this temperature is still above the magnetic ordering temperature of 7 K, this additional susceptibility results from non-Kitaev terms, as recently suggested by exact diagonalization (ED) calculations\cite{Rousochatzakis2019}. 

To further investigate the temperature and energy dependence of the non-Kitaev interactions, we consider the energy and temperature dependent colormap in Fig.~\ref{fig:Fig1}c. Here $\chi_{\delta} = \chi_{measured}-\chi_{QMC}$ is the difference between the measured Raman susceptibility and that of the pure Kitaev model (determined by the QMC calculation). The green color indicates the measured susceptibility is higher than the QMC results and the blue color indicates regions of very good overlap between the measurement and the calculation. The black dots suggest the temperature and energy boundaries where the system perfectly resembles the pure Kitaev QSL. Specifically, there is a large $\chi_{\delta}$ in the quarter circle area below 6 meV and 40 K, which can be explained as the region where non-Kitaev interactions become dominant in the response. The  deviation above 150 K and below 8 meV results from the QES induced by thermal fluctuations in the system, which is well known in frustrated magnets\cite{Han2012Nat,Glamazda2017,sandilands2015PRL,Yao2016CGT,Kim2019NatComm}. The high energy deviation ($>$12meV) is from the low energy tail of the phonon. Nonetheless, the low energy and temperature deviation from the pure Kitaev model is consistent with the calculated intensities of recent ED results for a model with only $\Gamma$ and Kitaev terms in the system (K-$\Gamma$ model)\cite{Rousochatzakis2019} with $-J_K/\Gamma = 5$. Furthermore, the ED results suggest enhanced response over that expected for the pure Kitaev limit for $\omega_{\Gamma} \sim 2.5 \Gamma$. As shown in our colormap, when the temperature is low, the disagreement occurs for $\omega_{\Gamma} < 6$ meV. based on K-$\Gamma$ model, this suggests $\Gamma \approx 2.4$ meV, where the Heisenberg interaction and terms beyond nearest neighbors are neglected. We note that regardless of the specific non-Kitaev terms, this can be interpreted as an upper bound on the ratio of Kitaev to non-Kitaev terms in this system. Furthermore, we find the best agreement for the pure Kitaev limit with $J_K = 10$ meV (see Supplemental Sup\_Fig.~4), consistent with our observed bandwidth of the continuum (Fig.~\ref{fig:Fig1}a) of 30 meV.\cite{sandilands2015PRL,knolle2014raman} The $\Gamma\approx 2.4$ meV we obtained here is also consistent with the results obtained from neutron scattering (2.5 meV)\cite{Winter2017NatComm}, from thermal Hall measurements (2.5 meV)\cite{Kasahara2017PRL}, and from THz measurements (2.4 meV)\cite{Wu2018PRB}. 

\begin{sloppypar}
Having established the size and extent of the non-Kitaev terms, we examine the statistics of the excitations in $\alpha$-RuCl$_{3}$ to see if they are truly fractional. As the statistics depends on both temperature and energy, one should make sure the system is in detailed balance\cite{devereaux2007RoM} and that laser heating is negligible, which was not quantitatively shown before. As discussed in the supplemental, the fermionic response written above is consistent with the fluctuation-dissipation theorem with the presence of time-reversal symmetry, requiring $I_{S}[\omega,T]/I_{aS}[\omega,T]=e^{\frac{\hbar\omega}{k_{B}T}}$\cite{devereaux2007RoM,Misochko1996}. Previously, the discrepancy between the prediction of the Bose factor and the measured intensity at low temperatures was attributed to fractional statistics\cite{Nasu2016,Glamazda2017}. However, these works did not exclude the possibility that laser induced heating kept the measured area at a fixed temperature, while the bulk was cooled. This is not unlikely, given the small specific heat and thermal conductivity of RuCl$_{3}$ at low temperatures\cite{Nasu2015PRB,Majumder2015PRB,Cao2016PRB,Hyeonsik2011PRB,Hirobe2017PRB}. Furthermore, as described in the supplemental, previous uses of the anti-Stokes responses were unreliable due to the low signal levels\cite{sandilands2015PRL}. Most importantly, unless the temperature is well known, it is difficult to directly compare with the theoretical prediction for fractional statistics. In our current work we have made substantial improvements to the thermal anchoring and collection efficiency to allow for much higher temperature resolution and lower Raman frequency. In this way, we can observe the spectra change between different temperatures and directly compare it with QMC. Most importantly, due to enhanced signals and lower probing frequencies, we have been able to collect anti-Stokes response at lower temperatures to ensure that laser induced heating is not an issue. Returning to the actual sample temperature, in Fig.~\ref{fig:Fig2}d, we compare the anti-Stokes intensity and Stokes intensity times a Boltzmann factor with the measured temperature. The excellent agreement between them reveals that there is nearly no heating in the laser spot and thus we can use the measured crystal temperature. Unlike previous studies\cite{sandilands2015PRL,Glamazda2017}, our new quantitative comparison between Stokes and anti-Stokes limits the possibility of laser heating to explain the low temperature upturn and confirms the sample is in detailed balance.
\end{sloppypar}

We explore the possibility that the Raman susceptibility results from purely fermionic excitions in Fig.~\ref{fig:Fig2}a. If the excitation is fractional, one expects $\text{Im}[\chi(\omega,T)]\propto JDos(\omega,T)*(1-2n_F(\omega/2,T))$. To cancel the constant term and focus only on the fermionic part, we show the difference of susceptibility: $\Delta Im(\chi[\omega,T\leq 150~\text{K}])= \text{Im}(\chi[\omega,T]) - \text{Im}(\chi[\omega,150~\text{K}])$. The utility of such an analysis is quite clear: the energy and temperature extent of the continuum can be directly observed - without contributions from high temperature QES fluctuations or phonons. To test the predicted fermionic response from fractional particles, we plot $\Delta n_F[\omega/2,T]=n_F[\omega/2,T]-n_F[\omega/2,150~\text{K}]$, as contour lines on top of the data. The agreement is quite good and is further confirmed in  Fig.~\ref{fig:Fig2}c via constant temperature cuts of the data shown in Fig.~\ref{fig:Fig2}a, along with the calculated $\Delta n_{F}[\omega/2,T]$. For Raman scattering, it is a particle  creation/absorption process, so the temperature and energy dependent of fractional particles is determined by occupation, which is described by Fermi function. Therefore, the good agreement between the data and Fermi functions  with  half  of  the  scattering  energy  provides  strong
evidence for the presence of pairs of fractional particles. We note this is done without any artificial subtraction of a bosonic background. This approach relies on a nearly temperature and energy independent $JDos[\omega,T]$, which is expected from numerical calculations for the Kitaev system at temperatures above the flux gap\cite{Nasu2015PRB}. This assumption appears to generally hold in our data, whose temperature an energy evolution are generally described by a Fermi function. Nonetheless at the lowest temperatures, there is some deviation of the data for energies above 6 meV. The origin of this discrepncy is not clear, but likely results from a the temperature and energy dependence of the $JDos[\omega,T]$. Additionally, we find poor agreement if the full scattering energy ($n_F[\omega,T]$) is used (not shown).  We additionally performed the same analysis on another honeycomb system Cr$_{2}$Ge$_{2}$Te$_{6}$ (Fig.~\ref{fig:Fig2}b), which was grown by established methods and which is ferromagnetic below 60 K with a similar Curie-Weiss temperature as $\alpha$-RuCl$_{3}$\cite{CGT2013,Yao2016CGT}. The behavior of Cr$_{2}$Ge$_{2}$Te$_{6}$ is the exact opposite of $\alpha$-RuCl$_{3}$, namely, $\Delta \text{Im}(\chi[\omega,T])$ is negative throughout the whole measured range and decreases upon cooling. 
 
\begin{figure}
    \centering
    \includegraphics[width=1\textwidth]{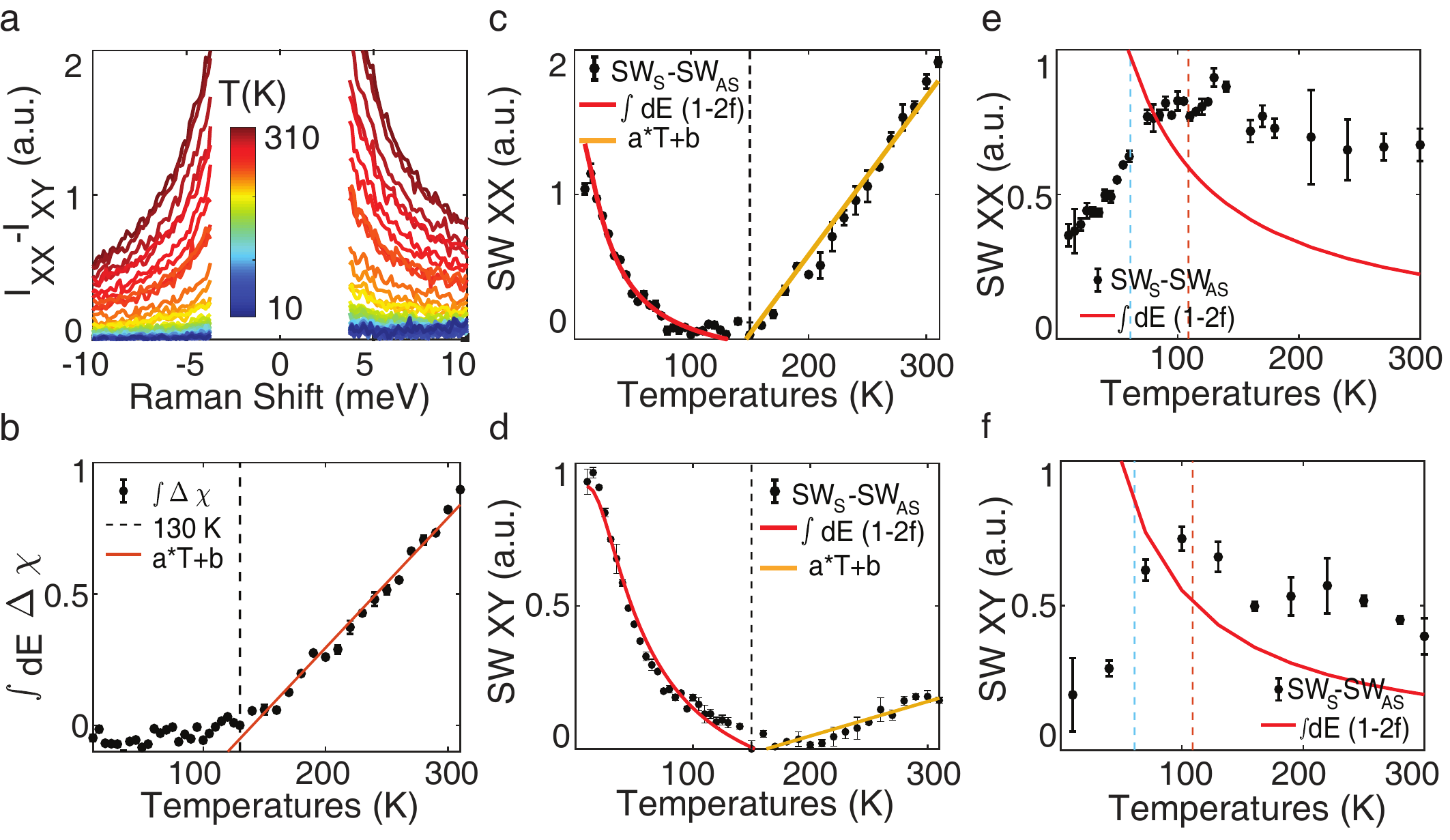}
    \caption{\textbf{Limit of Fermi statistics} 
     (\textbf{a}) The continuum in $\alpha$-RuCl$_{3}$ due to fractional particles is removed by taking the difference between XY and XX intensities. This confirms the continuum is consistent with predictions of the Kitaev model, and the high temperature response is from quasi-elastic scattering (i.e. Lorentzian times a Bose factor). (\textbf{b}) The integration of the Raman susceptibility(3meV - 8meV) with only the quasi-elastic scattering response, reveals a linear T behavior above 150 K and temperature independent behavior below. (\textbf{c} \& \textbf{d}) Integrated spectral weight(3meV - 8meV) of $\text{Im}[\chi(\omega,T)]$, reveals Fermi statistics in $\alpha$-RuCl$_{3}$ below $\approx$100 K (solid red line) in XX and XY  polarizations. Above 150 K the response is linear in temperature due to the quasi-elastic scattering (yellow lines). The spectral weight(3meV - 8meV) from Cr$_{2}$Ge$_{2}$Te$_{6}$ (\textbf{e} \& \textbf{f}) is enhanced up to $T_{C}$ (blue dashed line) but the temperature dependence above does not fit that expected for fermions (solid red line).}
    \label{fig:Fig3}
\end{figure}

Returning to Fig.~\ref{fig:Fig2}a, we have also drawn the boundary of the non-Kitaev contributions determined from the analysis in Fig.~\ref{fig:Fig1}c.We find the Fermi function matches the susceptibility very well, suggesting the Fermi statistics hold even when the agreement with the pure Kitaev model (see Fig. ~\ref{fig:Fig1}b) does not. However, given the relatively large size of the Kitaev term relative to the non-Kitaev contributions, this may not be surprising and suggests the excitations in $\alpha$-RuCl$_{3}$ are primarily fractionalized. Our analysis presented in Fig.~\ref{fig:Fig1}c and Fig.~\ref{fig:Fig2}a also reveal the crossover from spin liquid-like behavior (i.e. fractional continuum) to a standard paramagnet. Indeed, $\Delta \text{Im}(\chi[\omega,150~\text{K}\leq T\leq 200~\text{K}])$ is nearly constant, as expected for a paramagnet in this range. As discussed later, the response at higher temperatures is consistent with quasi-elastic scattering. We note that the exact  temperature  at  which  the  response  will  set  in, depends  on  the  energy  scale  at  which  it  is  measured. As such the  integrated  response  investigated  in  Fig.~\ref{fig:Fig3}C,  appears  to  have  a  higher  onset  temperature  for  the  QES  due  to  the inclusion of higher energy scales. Specifically, a Lorentzian at zero energy results from thermal fluctuations of the magnetism that confirm the magnetic specific heat is consistent with a standard paramagnet at high temperatures. Lastly, this analysis also provides new insights into the phonons overlapping the continuum. Specifically, consistent with previous works we also find the phonons have a low energy tail due to their coupling with the continuum (see Fig. ~\ref{fig:Fig1}c and supplemental Fig. ~4b). However, via our new comparison with the pure Kitaev limit, it is clear the influence of the two lowest energy phonons on the continuum is limited to $\approx 12\rightarrow 14meV$, and $\approx 15\rightarrow 19meV$, respectively. Interestingly, the response in these regions still follows the prediction of the Fermi function ( \ref{fig:Fig2}a), showing the mixing of the phonons with the fractional excitations does not significantly influence them. 
 
To ensure our approach is self-consistent it is worthwhile to also analyze the integrated Raman response, as done previously in $\alpha$-RuCl$_{3}$ and Li$_{2}$IrO$_{3}$.\cite{Nasu2016,sandilands2015PRL,Glamazda2017} Likewise, it is also crucial to find a reliable method to separate the QES response from the continuum such that it can be independently studied for further confirmation of the presence of Fermi statistics. This is now possible using both the polarization and Stokes minus anti-Stokes spectra ($\text{Im}[\chi[\omega,T]]$). Since the continuum has equal weight in both polarizations\cite{knolle2014raman,sandilands2015PRL} it can be removed via their difference: $\Delta I_{S/aS}[\omega,T]=I^{XX}_{S/aS}[\omega,T]-I^{XY}_{S/aS}[\omega,T]$. We also note the isotropic response of the continuum implies an isotropic Kitaev interaction.\cite{knolle2014raman,Nasu2016} As seen in Fig.~\ref{fig:Fig3}a, $\Delta I_{S/aS}[\omega,T]$ is consistent with thermal fluctuations (i.e. QES)\cite{wulferding2010PRB,Nakamura2015JPSJ,Kim2019NatComm}, namely a Lorentzian whose amplitude is given by the magnetic specific heat ($C_{m}[T]$) times the temperature and appropriately weighted Bose factors (i.e. greater Stokes than anti-Stokes intensity). We now calculate the QES amplitude via the spectral weight (SW) of the Raman susceptibility: $SW_{QES}[T]=\int \chi^{QES}_{XX}[\omega,T]-\chi^{QES}_{XY}[\omega,T]d\omega=\int dE\Delta \chi$. Here, the integration energy range is 3 to 8 meV. Consistent with direct fits of the $\Delta I_{S/aS}[\omega,T]$ (see supplemental) and robust to the limits of integration (as long as phonons are not included), we find $SW_{QES}[T]\propto T$ (see Fig.~\ref{fig:Fig3}b). This suggests the magnetic specific heat is nearly temperature independent, as expected for a classical paramagnet at high temperatures. Since the QES signal is nearly zero in $\chi[\omega,T<150~\text{K}]$, this confirms the Raman susceptibility (and not the intensity) naturally separates the QES from the continuum. Thus our new measurements reveal the energy and temperature range over which the excitations are fractional without contamination from other contributions.

Having isolated the QES and identified its temperature dependence, we can independently check the temperature bounds of the Fermi statistics. Specifically, we investigate the difference between the Stokes and anti-Stokes SW in a given polarization ($\Delta SW[T]=\int (I_{S}[\omega,T]-I_{aS}[\omega,T])d\omega$), which includes the integrated Fermi function from the fractional excitations and the QES contribution (see supplemental). As shown in Fig.~\ref{fig:Fig3}c \& d for two different polarizations, the integrated weight follows the expected response for pairs of fermionic excitations ($\int (1-2n_{f}(\omega/2,T))d\omega$) until $T\approx 150~\text{K}$ where it crosses over to a linear temperature dependence from the QES. The fermionic response is equal in both polarizations and covers the Kitaev ranges. Thus with just three parameters, one fixed by the lowest temperature, we fully explain the SW for all energy ranges, temperatures, and polarizations. To further confirm this, we tried the same analysis on our new Cr$_{2}$Ge$_{2}$Te$_{6}$ data. As shown in Fig.~\ref{fig:Fig3}e\& f, the difference between the Stokes and anti-Stokes of Cr$_{2}$Ge$_{2}$Te$_{6}$ cannot be fit with a Fermi function at all. Thus the results presented in Fig.~\ref{fig:Fig3}c,d provide a quantitative confirmation of the presence of fractional excitations up to high temperatures.


To conclude, our higher quality data and anti-Stokes spectra provide direct comparison of the Raman susceptibility energy and temperature dependence with QMC calculations. At higher temperatures and energies, these results are consistent with QMC calculations for the pure Kitaev limit. Consistent with ED calculations, the Raman susceptibility is enhanced over the Kitaev QSL only at low energies and temperatures due to additional non-Kitaev terms. Thus our results reveal the temperature and energy boundary of non-Kitaev interactions becoming dominant. Furthermore, via comparison of the measured Raman susceptibility and the Fermi function, the data provide concrete evidence that the magnetic excitations in $\alpha$-RuCl$_{3}$ are fractional and follow Fermi statistics. Interestingly, these fractional excitations follow Fermi statistics even in the ranges where non-Kitaev terms become dominant. It remains to be answered whether, and how, different non-Kitaev terms compete with each other in the low temperature and energy range. Nonetheless our approach enables a new means to extract the size and influence of non-frustrating terms in QSLs, and could be applied at finite magnetic field to confirm the fractional nature of excitations in the field induced QSL state of $\alpha$-RuCl$_{3}$.

\newpage
\begin{methods}

\noindent
\textbf{RuCl$_{3}$ crystal growth, handling and characterization.}
Single crystals of $\alpha$-RuCl$_{3}$ were prepared using high-temperature vapor-transport techniques from pure $\alpha$-RuCl$_{3}$ powder with no additional transport agent. Crystals grown by an identical method have been extensively characterized via bulk and neutron scattering techniques\cite{Cao2016PRB,Banerjee2017Science,Banerjee2018npj} revealing behavior consistent with what is expected for a relativistic Mott insulator with a large Kitaev interaction\cite{PLumb2014,Nasu2014,Sandilands2016PRB,Koitzsch_EELS_2016PRL,Zhou2016PRB,Yadav2016,Do2017Natphys,Hirobe2017PRB,Winter2017NatComm,Jansa2018NatPhys,Leahy2017PRL,Wellm2018,Figgis1966}. The crystals have been shown to consistently exhibit a single dominant magnetic phase at low temperature with a transition temperature $T_{N}\approx 7$ K, indicating high crystal quality with minimal stacking faults\cite{Cao2016PRB}. Care was taken in mounting the crystals to minimize the introduction of additional stacking faults, as evidenced by the high reproducibility of the spectra across different crystals and experimental setups. Characterization was consistent with previous studies\cite{PLumb2014,WangLoidl2017PRL,Zheng2017PRL,mashhadi2018NanoLett,Majumder2018PRL,Little2017PRL}.

\noindent
\textbf{Raman spectroscopy experiments.}
Since Raman scattering involves a photon in and photon out, it allows one to measure both the symmetry and energy change of an excitation. Furthermore, one can choose an energy and/or symmetry channel to separate the magnetic, electronic and lattice responses\cite{sandilands2015PRL,Nasu2016,devereaux2007RoM,wulferding2010PRB,Nakamura2015JPSJ,Yao2016CGT,Glamazda2017}. The majority of the Raman experiments were performed with a custom built, low temperature microscopy setup\cite{Tian2016RSI}. A 532 nm excitation laser, whose spot has a diameter of 2 $\mu m$, was used with the power limited to 30 $\mu$W to minimize sample heating while allowing for a strong enough signal. The sample was mounted by thermal epoxy onto a copper \emph{xyz} stage. At both room and base temperature the reported spectra were averaged from three spectra in the same environment to ensure reproducibility. The spectrometer had a 2400 g/mm grating, with an Andor CCD, providing a resolution of $\approx 1$ cm$^{-1}$. Dark counts are removed by subtracting data collected with the same integration time with the laser off. To minimize the effects of hysteresis from the crystal structural transition, data was taken by first cooling the crystal to base temperature, and once cooled to base temperature, spectra were acquired either every 5 or 10 K by directly heating to that temperature. The absence of hysteresis effects was confirmed by taking numerous spectra at the same temperature after different thermal cycles (100 K in the middle of the hysteresis region). In addition, recent studies of the Raman spectra of RuCl$_{3}$ suggest an effect of the surface structure upon exposure to air\cite{Zhou2018JPCS,mashhadi2018NanoLett}. To minimize this, crystals were freshly cleaved and immediately placed in vacuum within three minutes. Lastly, a recently developed wavelet based approach was employed to remove cosmic rays\cite{Tian2016RSI,Tian2016AS}.

\noindent
\textbf{Quantum Monte Carlo Calculations.} 
The Hamiltonian of the Kitaev model on the honeycomb lattice is given by
 \begin{equation}
  {\cal H}=-J_x\sum_{\means{jk}_x}S_j^x S_k^x-J_y\sum_{\means{jk}_y}S_j^y S_k^y-J_z\sum_{\means{jk}_z}S_j^z S_k^z,
  \label{eq:H_spin}
 \end{equation}
 where $\bm{S}_j$ represents an $S=1/2$ spin on site $j$, and $\means{J_K}_\gamma$ stands for a nearest-neighbor (NN) $\gamma(=x,y,z)$ bond shown in Fig.~1a.
In the calculation for the spectrum of the Raman scattering we adopt the Loudon-Fleury (LF) approach.
The LF operator for the Kitaev model is given by
\begin{equation}
 {\cal R}=\sum_{\means{ij}_\alpha}(\bm{\epsilon}_{\rm in}\cdot \bm{d}^\alpha)(\bm{\epsilon}_{\rm out}\cdot \bm{d}^\alpha)J_\alpha S_i^\alpha S_j^\alpha,\label{eq:LF}
\end{equation}
where $\epsilon_{\rm in}$ and $\epsilon_{\rm out}$ are the polarization vectors of the incoming and outgoing photons and $\bm{d}^\alpha$ is the vector connecting a NN $\alpha$ bond\cite{knolle2014raman,fleury1968scattering}.
Using this LF operator, the Raman spectrum is calculated as
\begin{equation}
 I(\omega)=\frac{1}{N}\int_{-\infty}^{\infty}dt e^{i\omega t}\means{{\cal R}(t){\cal R}},\label{eq:raman}
\end{equation}
where ${\cal R}(t)=e^{i{\cal H}t}{\cal R}e^{-i{\cal H}t}$ is the Heisenberg representation.
The temperature dependence of $I(\omega)$ is numerically evaluated using the Monte Carlo simulation in the Majorana fermion representation without any approximation\cite{Nasu2014}.
In the following we show the details of the calculation procedure\cite{Nasu2016}.

Using the Jordan-Wigner transformation, the Hamiltonian is mapped onto the Majorana fermion model as
\begin{equation}
   {\cal H}=\frac{iJ_x}{4}\sum_{(jj')_x}c_j c_k-\frac{iJ_y}{4}\sum_{(jj')_y}c_j c_k-\frac{iJ_z}{4}\sum_{(jj')_z} \eta_r c_j c_k,
\end{equation}
 where $(jj')_{\gamma}$ is the NN pair satisfying $j<j'$ on the $\gamma$ bond, and $\eta_{r}$ is a $Z_{2}$ conserved quantity defined on the $z$ bond ($r$ is the label for the bond), which takes $\pm 1$.
This Hamiltonian is simply written as
\begin{equation}
 {\cal H}=\frac{1}{2}\sum_{jk}A_{jk}(\{\eta_r\}) c_j c_k,
\end{equation}
using the Hermitian matrix $A_{jk}(\{\eta_{r}\})$ depending on the configuration of $\{\eta_{r}\}$.
The LF operator shown in Eq.~(\ref{eq:LF}) is also given by the bilinear form of the Majorana fermion:
\begin{equation}
 {\cal R}(\{\eta_r\})=\frac{1}{2}\sum_{jk}B_{jk}(\{\eta_r\}) c_j c_k,
\end{equation}
where $B(\{\eta_r\})$ is a Hermitian matrix.
To evaluate Eq.~(\ref{eq:raman}), we separate the sum over the states into $\{c_j\}$ and $\{\eta_r\}$ parts:
\begin{equation}
 I(\omega)= \frac{1}{Z}\sum_{\{\eta_r=\pm1\}}\bar{I}(\omega;\{\eta_r\})e^{-\beta F_f(\{\eta_r\})},
  \end{equation}
with
\begin{equation}
 \bar{I}(\omega;\{\eta_r\})=\frac{1}{Z_f(\{\eta_r\})}{\rm Tr}_{\{c_j\}}\left[\frac{1}{N}\int_{-\infty}^{\infty}dt e^{i\omega t}{\cal R}(t;\{\eta_r\}){\cal R}(\{\eta_r\})e^{-\beta {\cal H}(\{\eta_r\})}\right],\label{eq:raman_eta}
\end{equation}
where $Z=\sum_{\{\eta_r=\pm1\}}e^{-\beta F_f(\{\eta_r\})}$ and $Z_f(\{\eta_r\})=e^{-\beta F_f(\{\eta_r\})}={\rm Tr}_{\{c_j\}}e^{-\beta {\cal H}(\{\eta_r\})}$.
By applying Wick's theorem to Eq.~(\ref{eq:raman_eta}), we calculate the Raman spectrum at $\omega (\neq 0)$ for a given configuration $\{\eta_r\}$ as
\begin{align}
 \bar{I}(\omega;\{\eta_r\})=&\frac{1}{N}\sum_{\lambda\lambda'}\Bigl[
 2\pi|C_{\lambda\lambda'}|^2 f(\varepsilon_\lambda)[1-f(\varepsilon_{\lambda'})]\delta(\omega+\varepsilon_\lambda-\varepsilon_{\lambda'})\notag\\
 &+\pi|D_{\lambda\lambda'}|^2[1-f(\varepsilon_\lambda)][1-f(\varepsilon_{\lambda'})]\delta(\omega-\varepsilon_\lambda-\varepsilon_{\lambda'})\notag\\
 &+\pi|D_{\lambda\lambda'}|^2f(\varepsilon_\lambda)f(\varepsilon_{\lambda'})\delta(\omega+\varepsilon_\lambda+\varepsilon_{\lambda'}) \Bigr],
  \end{align}
where $f(\varepsilon)=1/(1+e^{\beta\varepsilon})$ is the Fermi distribution function with zero chemical potential, $\{\varepsilon_\lambda\}$ is the set of the positive eigenvalues of $A$ with the eigenvectors $\{\bm{u}_{\lambda}\}$, and the matrices $C$ and $D$ are given by $C_{\lambda\lambda'}=2 \bm{u}_{\lambda}^\dagger B \bm{u}_{\lambda'}$ and $D_{\lambda\lambda'}=2 \bm{u}_{\lambda}^\dagger B \bm{u}_{\lambda'}^*$.
In the Monte Carlo simulations, we generate a sequence of configurations of $\{\eta_r\}$ to reproduce the distribution of $e^{-\beta F_f(\{\eta_r\})}$, and hence the finite-temperature spectrum is simply computed as $I(\omega)=\means{\bar{I}(\omega;\{\eta_r\})}_{\rm MC}$ with $\means{\cdots}_{\rm MC}$ being the Monte Carlo average.

\noindent
\textbf{Correction for optical constants.}
According to the Beer-Lambert Law, the intensity of the laser decreases exponentially with the depth:
$I[z]=I_0 e^{-\alpha z}$, where $d$ is the depth and $\alpha$ is the attenuation constant, which is a function of laser frequency and dielectric constant of the material ($\alpha =\frac{\omega}{c}\text{Im}[\tilde{n}(\omega)] =-\frac{4\pi E[\omega_{0}]}{hc}k[\omega_{0}]$). Alternatively one can express this in terms of a penetration depth indicating the length scale relevant to absorption: $\delta=\frac{1}{\alpha}$. Applying this to our experiment, for a certain depth $d$, we find the incident laser intensity as a function of distance from the surface, $I_{in}[\omega_{0},z]=I_{0} e^{-\frac{4\pi E[\omega_{0}]}{hc}k[\omega_{0}]z}$. Here, $\omega_{0}$ is the frequency of the excitation laser, $I_{0}$ is the initial incoming laser power in front of the sample,and $\delta$($\approx$140 nm) is much shorter than the thickness of $\alpha$-RuCl$_{3}$ bulk crystal. 
To properly account for the temperature dependence of the optical constants on the measured Raman signal, it is crucial to account for these absorption losses. Specifically, the measured intensity is reduced by the absorption of the outgoing Raman photons, (i.e. $I_{out}[\omega,\omega_{0},z]=I_{in}[\omega_{0},z]e^{-\frac{4\pi E[\omega]}{hc}k[\omega]z}$) 
where $\omega$ is the frequency of the scattered light. Furthermore, one should also consider the probability of transmission at the surface of $\alpha$-RuCl$_{3}$ ($T[\omega]$), which also depends on the Raman light frequency. Applying the transmission rate to the Raman signal, we obtain the Raman intensity coming out of the sample at each point $I_{Raman}[\omega,\omega_{0},z] =I_{out}[\omega,\omega_{0},z] *T[\omega]$. Finally, one obtains the signal intensity by integrating the attenuated intensity of scattering point at each depth via $I_{corrected}[\omega_{0},\omega]=\int_{0}^{d_{max}} I_{Raman}[\omega,\omega_{0},z] dz $\cite{Tian2016RSI}.
All presented Raman data in this paper are corrected by this method using the previously published optical constants\cite{Sandilands2016PRB}. 
\end{methods}

\begin{addendum}
\item [Acknowledgements] We are grateful for numerous discussions with Natalia Perkins, Joshua Heath, Kevin Bedell and Ying Ran. The Raman experiments at 532 nm were performed by Y.W. with support from the National Science Foundation, Award No. DMR-1709987. G.B.O. assisted in the analysis with support from the U.S. Department of Energy (DOE), Office of Science, Office of Basic Energy Sciences under Award No. DE-SC0018675. Raman experiments performed at 720 nm (T.G. and J.Y.) were achieved by support from the National Science Foundation, Award No. ECCS 1509599. The crystal growth and characterization of $\alpha$-RuCl$_{3}$ (P.L.K. and D.M.) with , while A.B. and S.N. were supported by the US DOE Basic Energy Sciences Division of Scientific User Facilities. The work on Cr$_{2}$Ge$_{2}$Te$_{6}$ (H.J. and R.J.C.) at Princeton University is sponsored by an ARO MURI, grant W911NF1210461. The numerical simulations were performed by J.N. with support from Grants-in-Aid for Scientific Research (KAKENHI) (numbers JP15K13533, JP16H02206, JP16K17747, and JP18H04223). Parts of the numerical calculations were performed in the supercomputing systems in ISSP, the University of Tokyo.

\item[Competing Interests] The authors declare that they have no competing financial interests.

\item[Author Contributions] Y.W. performed the Raman experiments, with assistance from G.B.O., T.G., and J.Y. Analysis was done by Y.W., J.N., J.K., and G.B.O. The crystal growth and initial characterization were done by P.L., A.B., and D.M. The design of the experiments and conception of the study were achieved by K.S.B, S.N. and J.K. The theoretical calculations were performed by J.N., J.K. and Y.M.  

\item[Correspondence] Correspondence and requests for materials
should be addressed to K.S. Burch~(email: ks.burch@bc.edu).

\item[Data availability] All relevant data is available from the corresponding author
\end{addendum}

\bibliography{References}


\end{document}